# The Relationship Between Groundwater Nitrate Pollution and Crime in United States: Nitrate-Crime Hypothesis

Hossein Zamaninasab[1,*], Zohreh Bajelan[2]

1 Department of Biotechnology, College of Science, University of Tehran, Tehran, Iran. Email: zamani.nasab@ut.ac.ir, ORCID: https://orcid.org/0000-0002-7967-5945

2 Faculty of Chemistry and Petroleum Sciences, Shahid Beheshti University, Tehran, Iran. Email: z.bajelan@mail.sbu.ac.ir, ORCID: https://orcid.org/0009-0003-4657-6603

* Correspondence: zamani.nasab@ut.ac.ir

## Abstract

Groundwater is a crucial source of drinking water, but it is often contaminated with water-soluble pollutants that can pose significant health risks. Among these pollutants, nitrate has been extensively studied and found to be harmful to human health in high concentrations. However, the impact of nitrate pollution in groundwater on crime rates and social problems remains unexplored. This study aims to investigate the correlation between nitrate contamination in groundwater and crime rates in US states. The results suggest a potential link between nitrate pollution and increased crime rates, leading to the proposal of the "Nitrate-Crime Hypothesis." These findings highlight the importance of environmental factors in public health and safety policies and call for further research in this area.

**Keywords:** *Nitrate pollution; Groundwater; Crime rate; Nitrate-crime hypothesis*

# 1. Introduction

### *1.1. Access to Safe Drinking Water*

Access to safe drinking water has been one of the greatest challenges in the past few centuries. This issue has been particularly highlighted due to concerns about the physical or mental health of consumers; in fact, it has been shown that consuming contaminated drinking water with various chemical or microbial substances can lead to a wide range of diseases such as giardiasis, amoebiasis, ascariasis, multiple sclerosis, cancer, fetal damage in the mother's womb, stomach pain, etc. (Water Pollution Effects, 2006; Ward et al., 2018).

However, the possibility that consuming contaminated drinking water with significant levels of nitrates could lead to an increase in the incidence of social harm in a region has not yet been addressed. This possibility is not unlikely because it has been previously known that the chemical substances present in drinking water in an area can play a role in increasing crime and delinquency in the long run.

For example, the lead-crime hypothesis was first highlighted based on Needleman's research. The lead-crime hypothesis deals with the effect of increasing blood lead levels in childhood (e.g., due to drinking water via degraded lead pipes) on the occurrence of aggression and criminal behavior in later years (Needleman et al., 1996).

### *1.2. Groundwater as an Important Source of Drinking Water in the United States*

Groundwater is a freshwater source that is stored in rock pores and soil particles. Since ancient times, groundwater has been available to people for drinking in two ways: either it finds its way to the surface as springs or it is extracted from underground through well drilling. It is estimated that nearly 95 percent of the freshwater resources in the United States are allocated to groundwater sources (United States Environmental Protection Agency, 2023).

The underground water reserves in the United States are estimated to be at least 33 quadrillion gallons (equivalent to 1.25e14 cubic meters) (National Geographic Society, 1993), a significant portion of which is extracted annually.

It has been estimated that approximately 70% of this water is used for agricultural purposes (The Nature Conservancy, 2022). The National Groundwater Association estimates that the supply of drinking water for 4 out of 10 Americans depends on groundwater (National Groundwater Association, 2020).

Also, the United States Geological Survey estimates that up to 30 percent of surface water flow in the country is supplied through groundwater (The Nature Conservancy, 2022).

### *1.3. Nitrate Ion*

Nitrate is a molecular anion ion that consists of three oxygen atoms and one nitrogen atom per unit (Figure 1). Ionic compounds containing nitrate ions are common components in chemical fertilizers (Laue et al., 2006).

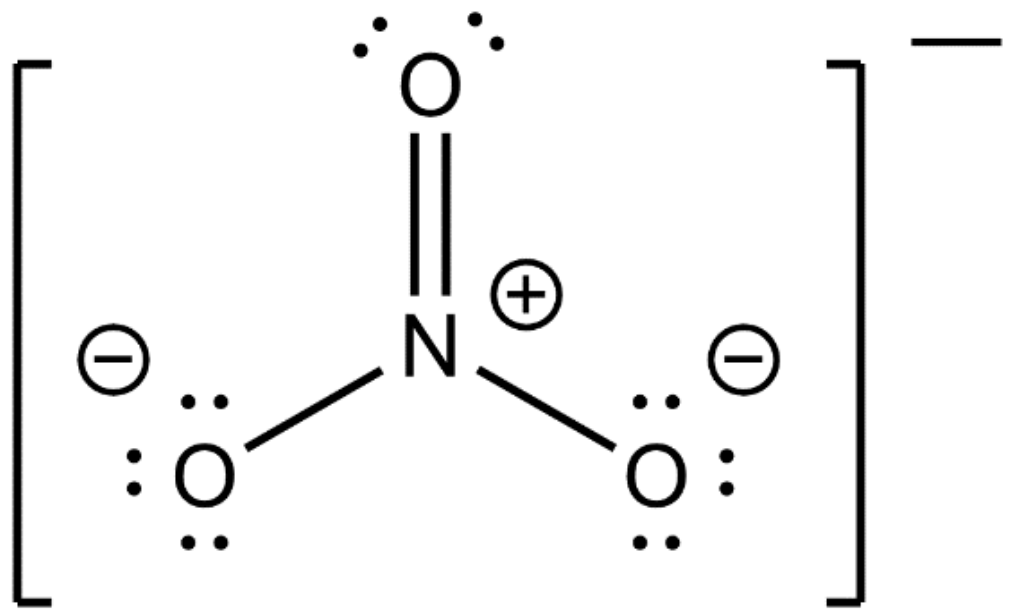

**Figure 1.** Lewis structure of nitrate ion

### *1.4. Nitrate Pollution Status in Drinking Water Sources in the United States*

Although some experts generally consider nitrate concentrations above 3 mg/L in groundwater as "contamination," (Madison & Brunett, 1985) the US Environmental Protection Agency (EPA) has set the Maximum Contaminant Level (MCL) for nitrate in water at 10 mg/L for nitrate as nitrogen. This figure has been set because higher nitrate concentrations can lead to infant methemoglobinemia, which is a dangerous and fatal syndrome. In other words, unfortunately, these regulations have not taken into account the numerous other adverse effects that drinking nitrate-contaminated water has on individual health (Messier et al., 2019).

This is particularly important because about 15 percent of the US population (over 43 million people) uses private wells as their drinking water source, whose quality and safety are not regulated by federal or state water laws (United States Geological Survey, 2019). It has been shown that 7% of these private wells across states and 22% of private wells in agricultural areas have excessive nitrate concentrations (Dubrovsky et al., 2010).

### *1.5. Social Problems in the United States*

The overall rate of criminality in the United States is high. The Global Organized Crime Index project has given the United States a criminality score of 5.5, ranking it 66th out of 193 countries in terms of criminality rate (The Organized Crime Index, 2023).

On average, there are over 460,000 victims of rape and sexual assault annually in the country (National Crime Victimization Survey, 2020). In 2020, women made up 89% of the victims, and men comprised 93% of the perpetrators (Federal Bureau of Investigation, 2021).

Additionally, according to data from the Department of Housing and Urban Development (HUD), in 2022, over 580,000 Americans will have experienced homelessness, averaging about 180 people per 100,000 population (with some states reaching up to 440 people per 100,000 population) (US Department of Housing and Urban Development, 2022).

This population is both more susceptible to committing crimes (Martell et al., 1995) and, at the same time, more likely to be victims of various forms of violence than others (Padgett & Struening, 1992). It is estimated that the annual cost of crime in the United States is around $4.7 to $5.8 trillion (Anderson, 2021).

### *1.6. Objectives*

This study aims to examine the relationship between nitrate concentrations in groundwater sources in U.S. states and the prevalence of various social problems in those states.

## 2. Materials and Methods

### *2.1. Study Design*

In this study, we first estimated the percentage of each state with groundwater nitrate concentrations >5 milligrams per liter for 48 states (data for Alaska and Hawaii were not available). Information on groundwater nitrates had previously been estimated using the USGS GWAVA-DW model (Nolan & Hitt, 2006) and reported by the U.S. Environmental Protection Agency based on data from 1991 to 2003 (United States Environmental Protection Agency, 2023) (see Appendix A for detailed information).

We then divided the states into two categories: "low-nitrate states," with less than 10% of their area having groundwater nitrate concentrations >5 milligrams per liter (n = 40), and "high-nitrate states" with more than 10% of their area having groundwater nitrate concentrations >5 milligrams per liter (n = 8). Finally, we examined how the average statistics of social problem indicators in low-nitrate states compared to those in high-nitrate states.

### *2.2. Study Indicators*

To better understand the social indicators under investigation, they were classified into five thematic categories. Table 1 presents information on these indicators and the relevant sources used to obtain them (see Appendix B for detailed information).

### *2.3. Software Used*

Results were categorized in tables using Microsoft® Excel® LTSC MSO (Version 2302, Build 16.0.16130.20186) 32-bit software. Apexcharts library version 3.36 and Paint.net 4.0 software were used to draw graphs. A heat map was drawn using ArcGIS Pro software (version 3.0.2).

**Table 1.** Information about the studied indices

| Category | Index | Description |
|---|---|---|
| General indicators | Crime index | This index represents the level of crime and violence in the U.S. states. A higher number indicates a higher level of crime (World Media Group, 2014). |
| | Violent crime | It is based on a report from the Federal Bureau of Investigation (FBI), which expresses the number of violent crimes in each state per 100,000 people. According to the FBI, violent crimes include murder, rape (revised definition), robbery, and aggravated assault (Federal Bureau of Investigation, 2014). |
| | Property crime | It shows the number of property crimes in each state per 100,000 people, based on the FBI's annual report (Federal Bureau of Investigation, 2014). |
| Theft and robbery | Robbery | It shows the robbery rate in each state per 100,000 people, based on the FBI's annual report (Federal Bureau of Investigation, 2014). |
| | Burglary | It shows the burglary rate in each state per 100,000 people, based on the FBI's annual report (Federal Bureau of Investigation, 2014). |
| | Larceny-theft | It shows the larceny-theft rate in each state per 100,000 people, based on the FBI's annual report (Federal Bureau of Investigation, 2014). |
| | Motor vehicle theft | It shows the rate of motor vehicle theft in each state per 100,000 people, based on the FBI's annual report (Federal Bureau of Investigation, 2014). |
| Homicide | Intentional homicide | It shows the murder rate in each state per 100,000 people, based on the FBI's annual report (Federal Bureau of Investigation, 2014). |
| | Murder and nonnegligent manslaughter | It shows the rate of murder and nonnegligent manslaughter in each state per 100,000 people, based on the FBI's annual report (Federal Bureau of Investigation, 2014). |
| Rape and assault | Aggravated assault | It shows the rate of aggravated assault in each state per 100,000 people, based on the FBI's annual report (Federal Bureau of Investigation, 2014). |
| | Rape (revised) | It shows the rate of rape (revised definition) in each state per 100,000 people, based on the FBI's annual report (Federal Bureau of Investigation, 2014). |
| | Rape (legacy) | It shows the rate of rape (legacy definition) in each state per 100,000 people, based on the FBI's annual report (Federal Bureau of Investigation, 2014). |
| Crime-related problems | Homelessness | It shows the rate of homelessness in each state per 100,000 people, based on the 2022 annual homelessness assessment report (US Department of Housing and Urban Development, 2022). |
| | Illiteracy | It shows the percentage of illiterate individuals in each state based on the report from the National Center for Education Statistics (NCES) (The National Center for Education Statistics, 2023). |
| | Unemployment | It shows the seasonally adjusted unemployment rate in each state (U.S. Bureau of Labor Statistics, 2023). |
| | Divorce rate | This shows the annual divorce rate per 1000 people in each state (National Vital Statistics System, 2021). |
| | Divorced percentage | It shows the percentage of Americans age 15 and older in each state who are divorced (U.S. Census Bureau, 2015). |

# 3. Results

## 3.1. Overview

On average, social problem indicators related to crime, such as the crime index, rates of various types of theft, homicide, aggravated assault, homelessness, illiteracy, and unemployment (14 out of 17 indicators), were worse in high-nitrate states. However, low-nitrate states had worse statistics for divorce and rape (Table 2).

**Table 2.** Comparison of the average of all indices in low-nitrate states with high-nitrate states

| Category | Index | Average value in low-nitrate states | Average value in high-nitrate states | Relative percentage* |
|---|---|---|---|---|
| General indicators | Crime index | 1534.10 | 1722.13 | + 12.3 % |
| | Violent crime | 332.71 | 392.11 | + 17.9 % |
| | Property crime | 2516.13 | 2642.64 | + 5.0 % |
| Theft and robbery | Robbery | 75.30 | 103.89 | + 38.0 % |
| | Burglary | 526.53 | 541.10 | + 2.8 % |
| | Larceny-theft | 1805.54 | 1882.40 | + 4.2 % |
| | Motor vehicle theft | 184.06 | 219.11 | + 19.1 % |
| Homicide | Intentional homicide | 5.95 | 6.71 | + 12.8 % |
| | Murder and nonnegligent manslaughter | 3.93 | 4.83 | + 22.9 % |
| Rape and assault | Aggravated assault | 214.28 | 247.13 | + 15.3 % |
| | Rape (revised) | 39.20 | 36.28 | - 7.4 % |
| | Rape (legacy) | 28.53 | 25.79 | - 9.6 % |
| Crime-related problems | Homelessness | 13.55 | 19.80 | + 46.1 % |
| | Illiteracy | 18.98 | 21.66 | + 14.1 % |
| | Unemployment | 3.15 | 3.33 | + 5.7 % |
| | Divorce rate | 2.78 | 2.20 | - 20.8 % |
| | Divorced percentage | 5.92 | 5.85 | - 1.2 % |

* Higher/lower percentage of high-nitrate states than low-nitrate states

## 3.2. Nitrate Pollution and General Crime Indicators

The average crime index was higher in high-nitrate states compared to low-nitrate states (with a ratio of 53:47). The rates of violent crime and property crime were also higher in high-nitrate states (with ratios of 54:46 and 51:49, respectively) (Figure 2).

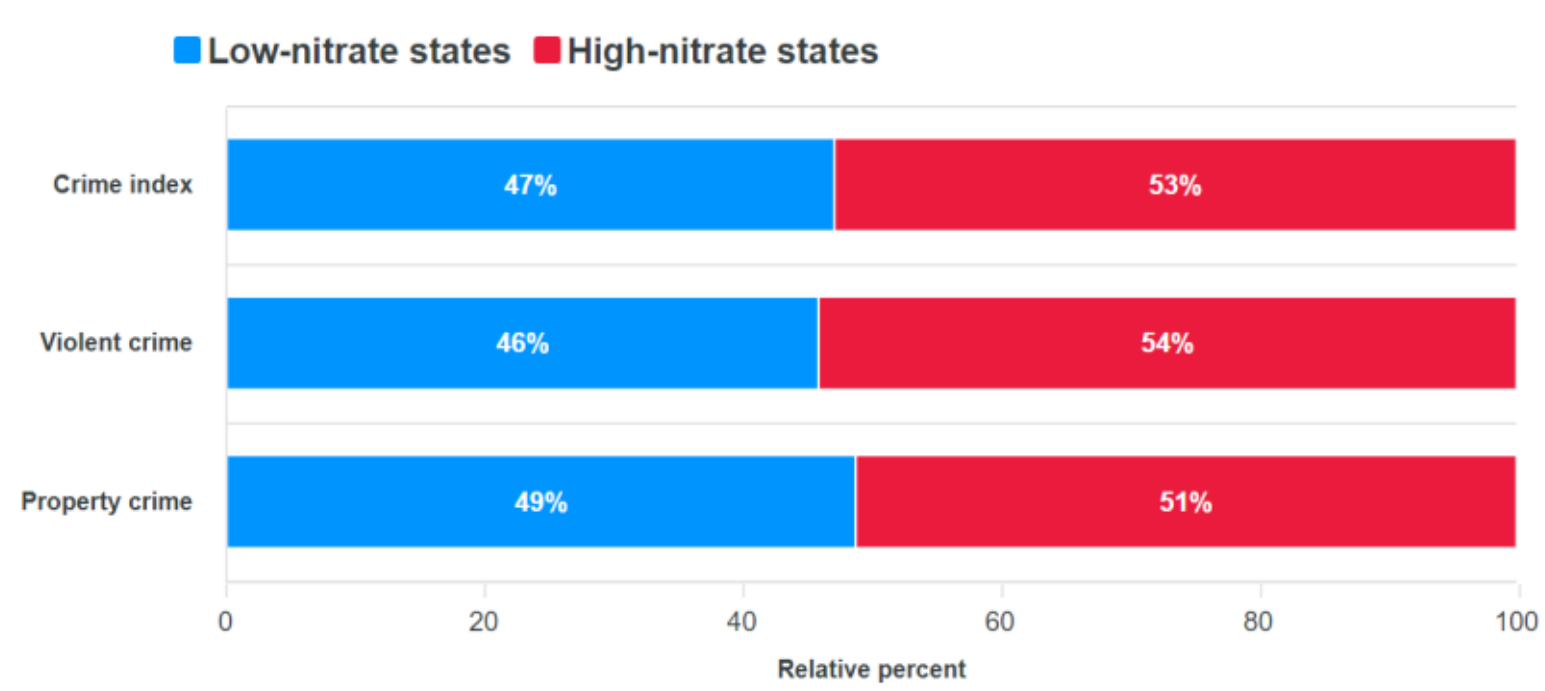

**Figure 2.** The average rate of some general crime indicators in the low-nitrate and high-nitrate states

### *3.3. Nitrate Pollution, Theft, and Robbery*

The average rates of robbery, burglary, larceny-theft, and motor vehicle theft per 100,000 population were higher in high-nitrate states (with ratios of 58:42, 51:49, 51:49, and 54:46, respectively) (Figure 3).

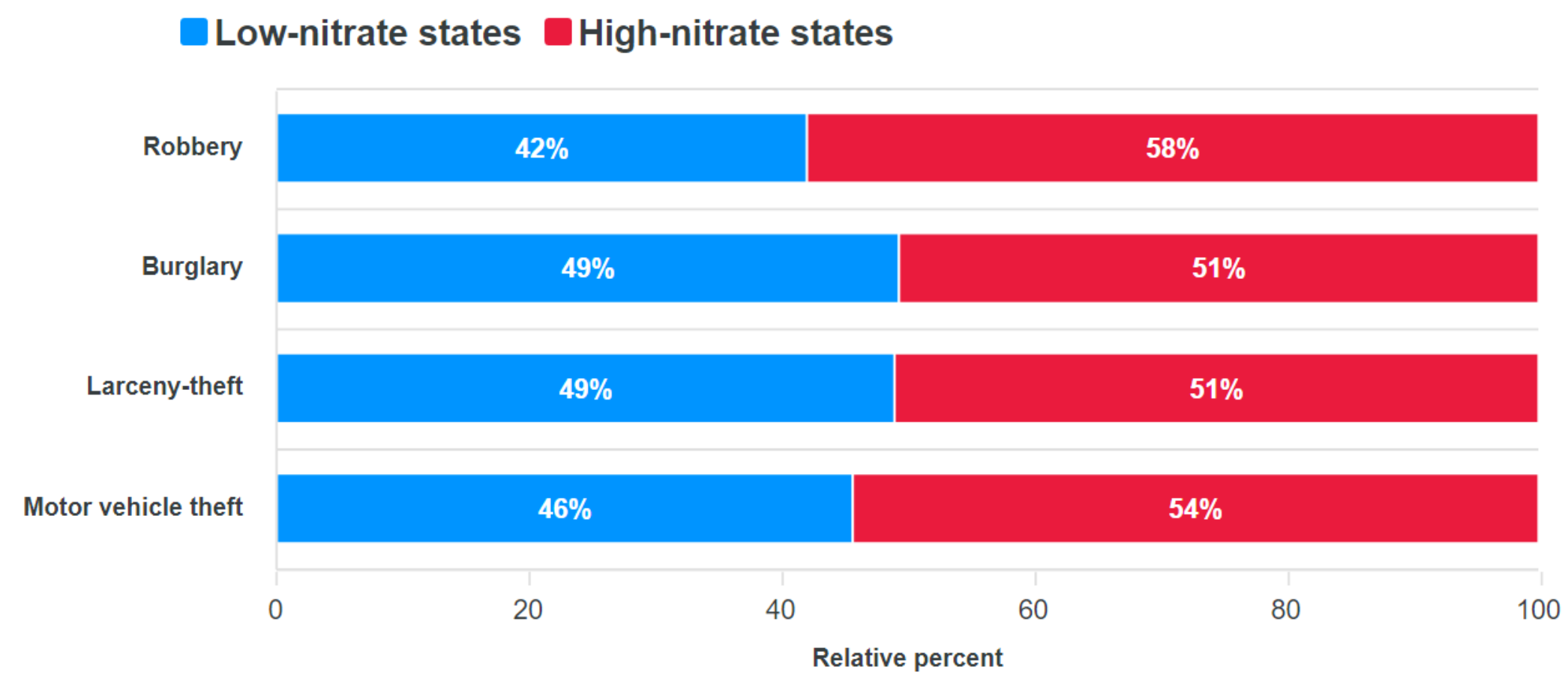

**Figure 3.** The average rate of theft and robbery in the low-nitrate and high-nitrate states

### *3.4. Nitrate Pollution and Homicide*

Both intentional homicide and nonnegligent manslaughter rates were higher in high-nitrate states (with ratios of 53:47 and 55:45, respectively) (Figure 4).

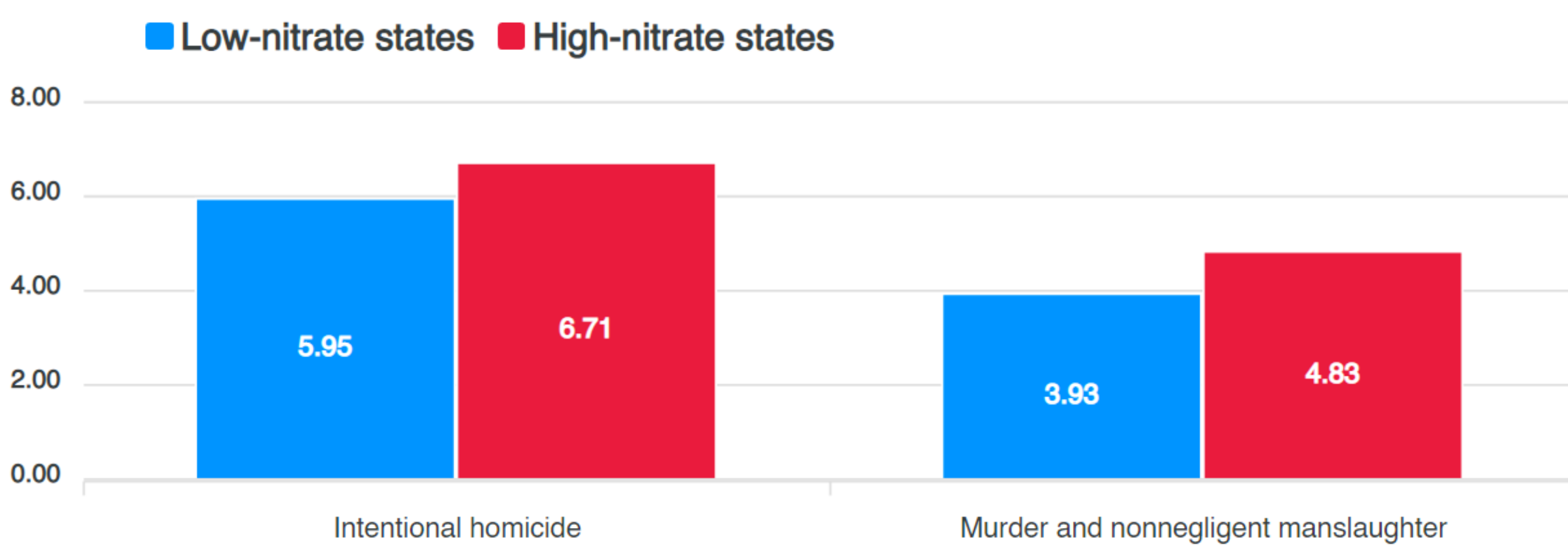

**Figure 4.** The average rate of homicide in the low-nitrate and high-nitrate states

### *3.5. Nitrate Pollution, Rape, and Assault*

The rate of aggravated assault was worse in high-nitrate states (with a ratio of 54:46), but the reported rape rate was worse in low-nitrate states (with ratios of 52:48 and 53:47, respectively) (Figure 5).

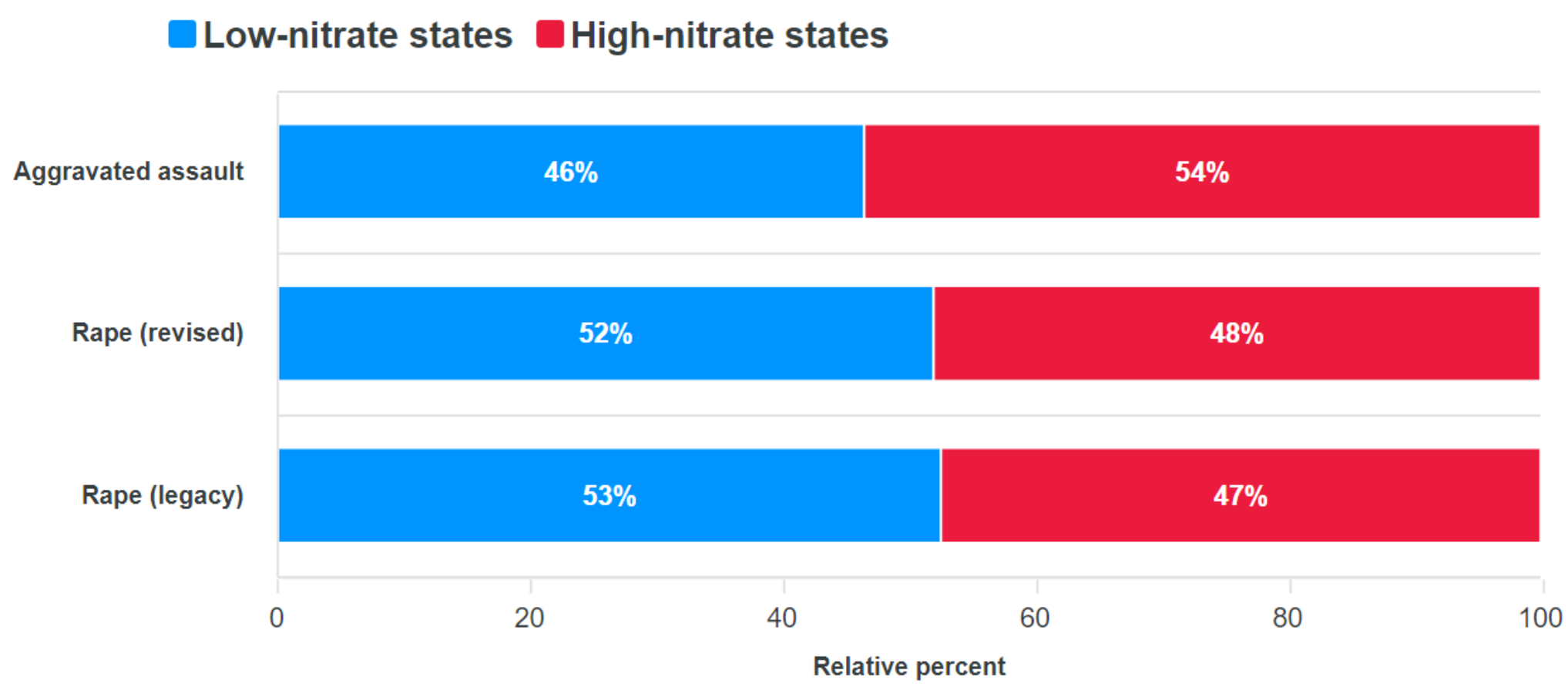

**Figure 5.** The average rate of aggravated assault and rape in the low-nitrate and high-nitrate states

### *3.6. Nitrate Pollution and Crime-related Problems*

Except for the annual divorce rate and percentage of divorced people (with ratios of 56:44 and 50:50, respectively), high-nitrate states had worse statistics for other social problems potentially related to crime, including homelessness, illiteracy, and unemployment (with ratios of 59:41, 53:47, and 51:49, respectively) (Figure 6).

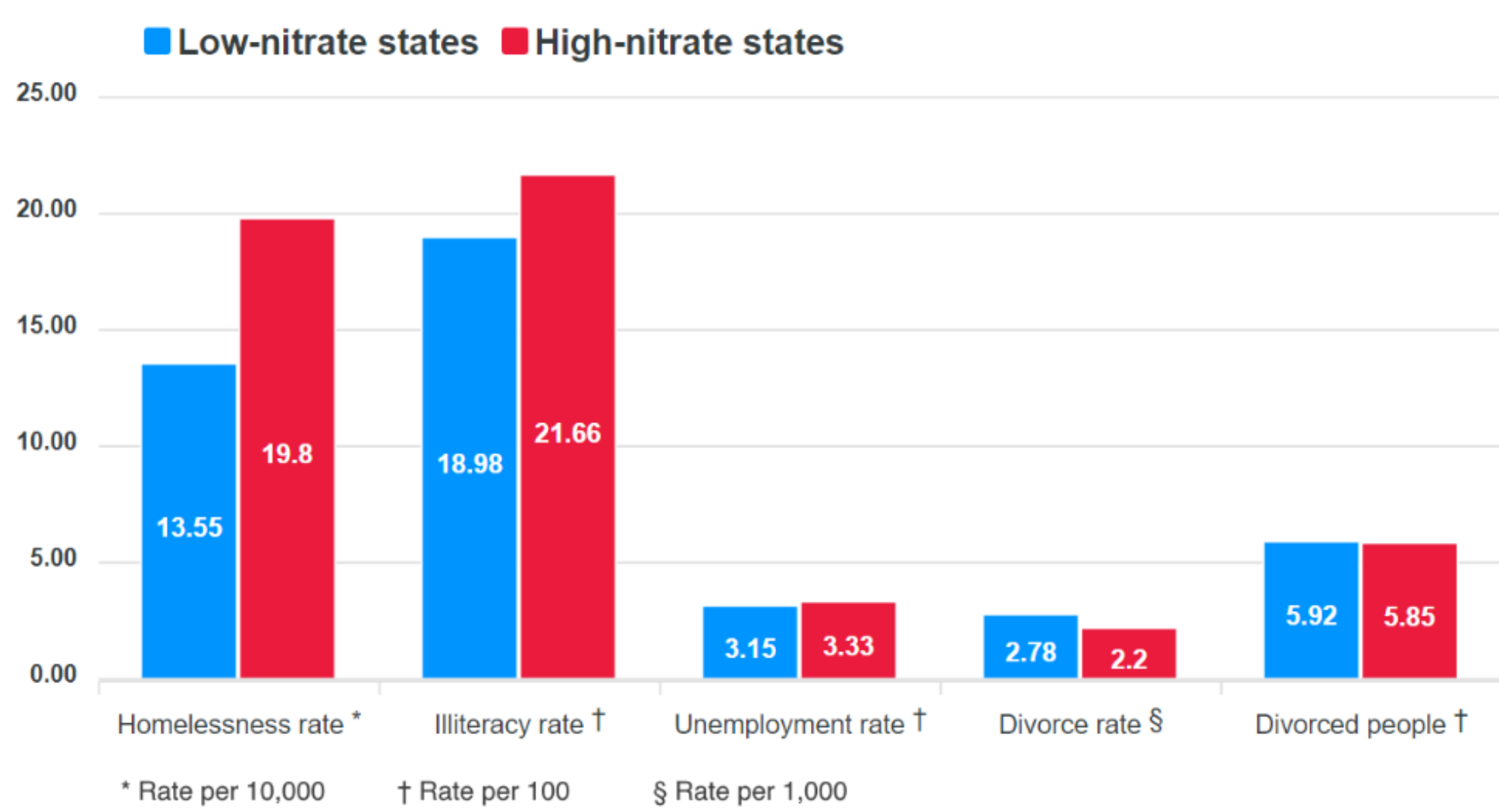

**Figure 6.** The average rate of crime-related problems in the low-nitrate and high-nitrate states

## 4. Discussion

### *4.1. Relationship between Nitrate Pollution and Crime*

As observed in the results section, almost all figures related to crime rates in a state were higher in high-nitrate states. This finding could be of interest to policymakers in the field of reducing social problems.

However, this study is just a preliminary step since many questions remain unanswered due to the lack of statistical analysis:

1. Are we facing a correlation or a causal relationship?
2. Are one or two high-nitrate or low-nitrate states responsible for the observed differences in crime rates, and ignoring them would make the difference disappear?
3. What would happen if we also considered the percentage of people in each state who obtain their drinking water from unregulated private wells?

### *4.2. Higher Rates of Divorce and Rape in Low-Nitrate States*

One exception was the reported rape rate, which was worse in low-nitrate states. It is not clear why the rape rate did not follow the same pattern as other indicators studied. However, one possibility is that since the figures reported by the FBI are related to only reported (and not unreported) rapes, low-nitrate states with lower illiteracy rates and from higher cultural classes may report any type of rape, no matter how small it is, while relatively illiterate people in high-nitrate states only report obvious cases of rape.

Also, the annual divorce rate and percentage of divorced individuals in the population of a state were worse in low-nitrate states. This issue should also be carefully considered; for example, we should pay attention to the issue of distinguishing unilateral and consensual divorces, as unilateral divorce (not consensual divorce, which is accompanied by mutual agreement) has the most impact on increasing crime and social problems in an area (Delpiano & Giolito, 2008). However, in this study, the overall prevalence of divorces was investigated, not unilateral and non-consensual divorces.

## 5. Conclusion

The results of the present study suggest a potential correlation between nitrate pollution and crime rates in U.S. states. However, further analysis is needed to determine if this relationship is causal and to identify any potential confounding factors. It is also worth noting that low-nitrate states had higher rates of reported rape and divorce, indicating that other factors may be at play in these areas. We propose the "Nitrate-Crime Hypothesis", in which the nitrate concentration in drinking water sources can be used as an indicator to increase the risk of a wide range of crimes in an area. Additional research is needed to confirm, reject, or modify this hypothesis.

# Appendix A

## Ranking of U.S. states based on estimated % of state area with groundwater nitrate concentrations > 5 mg/L

**Table A1.** U.S. states by estimated % of area with groundwater nitrate concentrations more or less than 5 mg/L (Nolan & Hitt, 2006)

| Rank | State | Estimated % of state area with groundwater nitrate concentrations > 5 mg/L | Rank | State | Estimated % of state area with groundwater nitrate concentrations > 5 mg/L |
|---|---|---|---|---|---|
| **1** | Delaware | 53% | **26** | Maine | 4% |
| **2** | Maryland | 28% | **27** | Minnesota | 4% |
| **3** | Nebraska | 17% | **28** | Arkansas | 3% |
| **4** | Rhode Island | 16% | **29** | New York | 3% |
| **5** | Louisiana | 15% | **30** | Oregon | 3% |
| **6** | Arizona | 12% | **31** | South Carolina | 3% |
| **7** | Massachusetts | 12% | **32** | Georgia | 2% |
| **8** | California | 10% | **33** | New Mexico | 2% |
| **9** | Florida | 9% | **34** | Utah | 2% |
| **10** | New Jersey | 9% | **35** | Virginia | 2% |
| **11** | North Carolina | 9% | **36** | West Virginia | 2% |
| **12** | Kansas | 8% | **37** | Wisconsin | 2% |
| **13** | Washington | 8% | **38** | Wyoming | 2% |
| **14** | Pennsylvania | 7% | **39** | Alabama | 1% |
| **15** | Vermont | 7% | **40** | Mississippi | 1% |
| **16** | Connecticut | 6% | **41** | Montana | 1% |
| **17** | Illinois | 6% | **42** | Nevada | 1% |
| **18** | Michigan | 6% | **43** | North Dakota | 1% |
| **19** | Texas | 6% | **44** | Oklahoma | 1% |
| **20** | Idaho | 5% | **45** | Tennessee | 1% |
| **21** | Ohio | 5% | **46** | Missouri | 0% |
| **22** | Colorado | 4% | **47** | New Hampshire | 0% |
| **23** | Indiana | 4% | **48** | South Dakota | 0% |
| **24** | Iowa | 4% | **-** | Alaska | No data |
| **25** | Kentucky | 4% | **-** | Hawaii | No data |

# Appendix B

## Ranking of American states based on different social problems

**Table B1.** U.S. states by some general crime indicators (World Media Group, 2014)

| Rank | States* | Crime index | Violent crime | Property crime |
|---|---|---|---|---|
| **1** | South Carolina | 2253 | 497.7 | 3460.3 |
| **2** | New Mexico | 2241 | 597.4 | 3542.3 |
| **3** | Tennessee | 2194 | 608.4 | 3060.6 |
| **4** | Nevada | 2180 | 635.6 | 2625.4 |
| **5** | Louisiana | 2167 | 514.7 | 3458.8 |
| **6** | Florida | 2095 | 540.5 | 3415.5 |
| **7** | Arkansas | 2065 | 480.1 | 3338 |
| **8** | Arizona | 2025 | 399.9 | 3197.5 |
| **9** | Delaware | 1974 | 489.1 | 2982 |
| **10** | Texas | 1962 | 405.9 | 3019.4 |
| **11** | Alabama | 1941 | 427.4 | 3177.6 |
| **12** | Oklahoma | 1938 | 406 | 2990.7 |
| **13** | Georgia | 1911 | 377.3 | 3281.2 |
| **14** | Maryland | 1884 | 446.1 | 2507.5 |
| **15** | North Carolina | 1869 | 329.5 | 2873.1 |
| **16** | Missouri | 1858 | 442.9 | 2906.5 |
| **17** | Washington | 1855 | 285.2 | 3706.1 |
| **18** | Michigan | 1757 | 427.3 | 2043.9 |
| **19** | California | 1701 | 396.1 | 2441.1 |
| **20** | Ohio | 1691 | 284.9 | 2799.1 |
| **21** | Kansas | 1670 | 348.6 | 2735.2 |
| **22** | Illinois | 1616 | 370 | 2075.9 |
| **23** | Indiana | 1599 | 365.3 | 2649.4 |
| **24** | Mississippi | 1584 | 278.5 | 2921.2 |
| **25** | Colorado | 1526 | 309.1 | 2530.1 |
| **26** | Oregon | 1497 | 232.3 | 2879 |
| **27** | Utah | 1405 | 215.6 | 2878.5 |
| **28** | Nebraska | 1390 | 280.4 | 2523.5 |
| **29** | Massachusetts | 1368 | 391.4 | 1857.1 |
| **30** | Pennsylvania | 1335 | 314.1 | 1931.7 |
| **31** | Kentucky | 1305 | 211.6 | 2246.9 |
| **32** | Minnesota | 1298 | 229.1 | 2297.5 |
| **33** | Rhode Island | 1268 | 219.2 | 2173.6 |
| **34** | West Virginia | 1235 | 302 | 2034.7 |
| **35** | Montana | 1230 | 323.7 | 2472.9 |
| **36** | Wisconsin | 1227 | 290.3 | 2088.3 |
| **37** | Iowa | 1203 | 273.5 | 2093.8 |
| **38** | New York | 1194 | 381.8 | 1718.2 |
| **39** | New Jersey | 1180 | 261.2 | 1734.1 |
| **40** | Connecticut | 1171 | 236.9 | 1920.4 |
| **41** | Wyoming | 1136 | 195.5 | 1964.7 |
| **42** | Virginia | 1125 | 196.2 | 1930.3 |
| **43** | Idaho | 1068 | 212.2 | 1854.8 |
| **44** | South Dakota | 1042 | 326.5 | 1863.9 |
| **45** | Maine | 1017 | 127.8 | 1986.4 |
| **46** | North Dakota | 985 | 265.1 | 2110.3 |
| **47** | New Hampshire | 956 | 196.1 | 1962.7 |
| **48** | Vermont | 950 | 99.3 | 1524.4 |

* States are sorted by crime index.

**Table B2.** U.S. states by theft and robbery rate (Federal Bureau of Investigation, 2014)

| Rank | States* | Robbery | Burglary | Larceny-theft | Motor vehicle theft |
|---|---|---|---|---|---|
| 1 | Nevada | 209.7 | 772.3 | 1494.3 | 358.7 |
| 2 | Maryland | 159.7 | 468.7 | 1819.6 | 219.2 |
| 3 | Delaware | 135.6 | 616.5 | 2230.1 | 135.4 |
| 4 | California | 125.5 | 522.3 | 1527.4 | 391.3 |
| 5 | Florida | 125.2 | 719.9 | 2481.5 | 214 |
| 6 | Georgia | 123 | 756.9 | 2258.4 | 266 |
| 7 | Louisiana | 122.5 | 824.5 | 2421.6 | 212.7 |
| 8 | New York | 121.8 | 257.2 | 1381.4 | 79.7 |
| 9 | Illinois | 118.8 | 388.2 | 1552.2 | 135.5 |
| 10 | New Jersey | 117.5 | 354.8 | 1248.3 | 131 |
| 11 | Texas | 115.7 | 627.8 | 2137.3 | 254.3 |
| 12 | Tennessee | 110.9 | 712.2 | 2156 | 192.4 |
| 13 | Ohio | 110 | 680 | 1963.6 | 155.4 |
| 14 | Pennsylvania | 105.8 | 357.5 | 1472.2 | 102 |
| 15 | Indiana | 104.5 | 559.3 | 1880 | 210.1 |
| 16 | New Mexico | 100 | 887.3 | 2353.4 | 301.6 |
| 17 | Alabama | 96.9 | 819 | 2149.5 | 209.1 |
| 18 | Arizona | 92.8 | 647.1 | 2289.1 | 261.3 |
| 19 | Missouri | 92.2 | 581.5 | 2055.3 | 269.8 |
| 20 | Massachusetts | 89.5 | 370.1 | 1364.5 | 122.5 |
| 21 | Wisconsin | 88 | 368.5 | 1547.6 | 172.3 |
| 22 | Connecticut | 87.8 | 332.4 | 1418.1 | 169.9 |
| 23 | North Carolina | 84.6 | 798.2 | 1937.8 | 137.1 |
| 24 | South Carolina | 82.7 | 759.9 | 2433.4 | 267 |
| 25 | Mississippi | 81.2 | 813.3 | 1956.9 | 150.9 |
| 26 | Michigan | 80.9 | 445.9 | 1384.5 | 213.5 |
| 27 | Washington | 79.9 | 783 | 2489.1 | 434 |
| 28 | Oklahoma | 78.6 | 760.9 | 1956.9 | 272.9 |
| 29 | Kentucky | 75.6 | 526.7 | 1577.1 | 143.2 |
| 30 | Arkansas | 69.1 | 835.7 | 2313.5 | 188.8 |
| 31 | Minnesota | 67.6 | 380.7 | 1763.5 | 153.3 |
| 32 | Colorado | 56.7 | 438.2 | 1857.1 | 234.8 |
| 33 | Nebraska | 55.4 | 422.5 | 1864.1 | 236.8 |
| 34 | Oregon | 52.7 | 434 | 2204.6 | 240.5 |
| 35 | Virginia | 51.5 | 277.7 | 1560.5 | 92.1 |
| 36 | Rhode Island | 50.1 | 457.1 | 1542.8 | 173.7 |
| 37 | Kansas | 46.9 | 545 | 1952.4 | 237.8 |
| 38 | Utah | 44.6 | 391.4 | 2239.1 | 248 |
| 39 | New Hampshire | 40.5 | 313.7 | 1584.4 | 64.6 |
| 40 | West Virginia | 35.2 | 484.9 | 1447.3 | 102.5 |
| 41 | Iowa | 33.6 | 464.4 | 1495.8 | 133.6 |
| 42 | North Dakota | 23.4 | 366.1 | 1539.5 | 204.7 |
| 43 | South Dakota | 23.4 | 330.3 | 1415.5 | 118 |
| 44 | Maine | 22.9 | 378.2 | 1548.2 | 60.1 |
| 45 | Montana | 19.8 | 351.2 | 1922.1 | 199.6 |
| 46 | Idaho | 12.5 | 393.3 | 1359.9 | 101.6 |
| 47 | Vermont | 11.2 | 324.6 | 1160.8 | 38.9 |
| 48 | Wyoming | 9.1 | 289.1 | 1572.4 | 103.2 |

* States are sorted by robbery rate.

**Table B3.** U.S. states by the rate of homelessness (US Department of Housing and Urban Development, 2022)

| Rank | States | Homelessness | Rank | States | Homelessness |
|---|---|---|---|---|---|
| **1** | California | 43.95 | **25** | Georgia | 9.79 |
| **2** | Vermont | 42.96 | **26** | Pennsylvania | 9.78 |
| **3** | Oregon | 42.35 | **27** | Missouri | 9.7 |
| **4** | New York | 37.7 | **28** | New Jersey | 9.45 |
| **5** | Washington | 32.38 | **29** | Oklahoma | 9.34 |
| **6** | Maine | 31.84 | **30** | Ohio | 9.06 |
| **7** | Nevada | 23.97 | **31** | Kentucky | 8.83 |
| **8** | Delaware | 23.26 | **32** | North Carolina | 8.77 |
| **9** | Massachusetts | 22.21 | **33** | Maryland | 8.68 |
| **10** | Arizona | 18.42 | **34** | Michigan | 8.18 |
| **11** | Colorado | 17.8 | **35** | Kansas | 8.16 |
| **12** | Louisiana | 16.06 | **36** | Texas | 8.14 |
| **13** | South Dakota | 15.27 | **37** | Wisconsin | 8.1 |
| **14** | Tennessee | 14.99 | **38** | Connecticut | 8.08 |
| **15** | Rhode Island | 14.42 | **39** | Arkansas | 8.07 |
| **16** | Montana | 14.12 | **40** | Indiana | 7.97 |
| **17** | Minnesota | 13.85 | **41** | North Dakota | 7.83 |
| **18** | New Mexico | 12.11 | **42** | West Virginia | 7.75 |
| **19** | Florida | 11.67 | **43** | Iowa | 7.56 |
| **20** | New Hampshire | 11.5 | **44** | Virginia | 7.52 |
| **21** | Nebraska | 11.41 | **45** | Alabama | 7.39 |
| **22** | Wyoming | 11.15 | **46** | Illinois | 7.32 |
| **23** | Utah | 10.52 | **47** | South Carolina | 6.83 |
| **24** | Idaho | 10.3 | **48** | Mississippi | 4.07 |

**Table B4.** U.S. states by the rate of illiteracy (The National Center for Education Statistics, 2023)

| Rank | States | Illiteracy | Rank | States | Illiteracy |
|---|---|---|---|---|---|
| **1** | New Mexico | 29.1 | **25** | Virginia | 18.8 |
| **2** | California | 28.4 | **26** | Indiana | 18.7 |
| **3** | Texas | 28.2 | **27** | Pennsylvania | 18.1 |
| **4** | Mississippi | 28 | **28** | Ohio | 17.7 |
| **5** | Louisiana | 27.1 | **29** | Michigan | 17.6 |
| **6** | Nevada | 25.3 | **30** | Massachusetts | 17.3 |
| **7** | New York | 24.4 | **31** | Connecticut | 17.2 |
| **8** | Alabama | 23.9 | **32** | Kansas | 16.9 |
| **9** | Florida | 23.7 | **33** | Oregon | 16.8 |
| **10** | Georgia | 23.6 | **34** | Colorado | 16.6 |
| **11** | Arizona | 23.4 | **35** | Idaho | 16.4 |
| **12** | Arkansas | 23.1 | **36** | Nebraska | 16.4 |
| **13** | South Carolina | 22.4 | **37** | Washington | 16.1 |
| **14** | Kentucky | 21.9 | **38** | Wisconsin | 15.3 |
| **15** | Tennessee | 21.7 | **39** | Iowa | 14.9 |
| **16** | North Carolina | 21.3 | **40** | South Dakota | 14.9 |
| **17** | West Virginia | 20.9 | **41** | Utah | 14.5 |
| **18** | New Jersey | 20.7 | **42** | Wyoming | 13.6 |
| **19** | Illinois | 20.4 | **43** | Maine | 13.4 |
| **20** | Rhode Island | 20.4 | **44** | North Dakota | 13.4 |
| **21** | Delaware | 20.3 | **45** | Minnesota | 13.1 |
| **22** | Oklahoma | 20.1 | **46** | Montana | 13.1 |
| **23** | Maryland | 20 | **47** | Vermont | 12.8 |
| **24** | Missouri | 18.9 | **48** | New Hampshire | 11.5 |

**Table B5.** U.S. states by the rate of unemployment (U.S. Bureau of Labor Statistics, 2023)

| Rank | States | Unemployment | Rank | States | Unemployment |
|---|---|---|---|---|---|
| **1** | Nevada | 5.4 | **25** | South Carolina | 3.1 |
| **2** | California | 4.5 | **26** | Virginia | 3.1 |
| **3** | Delaware | 4.3 | **27** | Indiana | 3 |
| **4** | Washington | 4.3 | **28** | Rhode Island | 3 |
| **5** | Illinois | 4.2 | **29** | Kansas | 2.9 |
| **6** | Pennsylvania | 4.1 | **30** | Oklahoma | 2.9 |
| **7** | New York | 4 | **31** | Arkansas | 2.8 |
| **8** | Oregon | 4 | **32** | Colorado | 2.8 |
| **9** | Texas | 4 | **33** | Minnesota | 2.8 |
| **10** | Connecticut | 3.8 | **34** | Iowa | 2.7 |
| **11** | Michigan | 3.8 | **35** | Florida | 2.6 |
| **12** | Kentucky | 3.7 | **36** | Idaho | 2.6 |
| **13** | Ohio | 3.7 | **37** | Maryland | 2.5 |
| **14** | Louisiana | 3.6 | **38** | Missouri | 2.5 |
| **15** | New Jersey | 3.5 | **39** | Maine | 2.4 |
| **16** | New Mexico | 3.5 | **40** | Vermont | 2.4 |
| **17** | Wyoming | 3.5 | **41** | Wisconsin | 2.4 |
| **18** | Arizona | 3.4 | **42** | Montana | 2.3 |
| **19** | Mississippi | 3.4 | **43** | Utah | 2.3 |
| **20** | North Carolina | 3.4 | **44** | Alabama | 2.2 |
| **21** | Massachusetts | 3.3 | **45** | New Hampshire | 2.1 |
| **22** | Tennessee | 3.3 | **46** | North Dakota | 2.1 |
| **23** | West Virginia | 3.3 | **47** | Nebraska | 2 |
| **24** | Georgia | 3.1 | **48** | South Dakota | 1.9 |

**Table B6.** U.S. states by the rate of homicide (Federal Bureau of Investigation, 2014)

| Rank | States* | Intentional homicide | Murder and nonnegligent manslaughter | Rank | States | Intentional homicide | Murder and nonnegligent manslaughter |
|---|---|---|---|---|---|---|---|
| **1** | Louisiana | 15.8 | 10.3 | **25** | Nevada | 5.7 | 6 |
| **2** | Missouri | 11.8 | 6.6 | **26** | California | 5.6 | 4.4 |
| **3** | Mississippi | 10.6 | 8.6 | **27** | Wisconsin | 5.3 | 2.9 |
| **4** | Arkansas | 10.6 | 5.6 | **28** | Colorado | 5.1 | 2.8 |
| **5** | South Carolina | 10.5 | 6.4 | **29** | Montana | 5 | 3.6 |
| **6** | Alabama | 9.6 | 5.7 | **30** | South Dakota | 4.5 | 2.3 |
| **7** | Tennessee | 9.6 | 5.7 | **31** | New York | 4.2 | 3.1 |
| **8** | Maryland | 9.1 | 6.1 | **32** | North Dakota | 4.2 | 3 |
| **9** | Illinois | 9.1 | 5.3 | **33** | Washington | 3.9 | 2.5 |
| **10** | Georgia | 8.8 | 5.7 | **34** | Connecticut | 3.9 | 2.4 |
| **11** | North Carolina | 8 | 5.1 | **35** | New Jersey | 3.7 | 3.9 |
| **12** | Pennsylvania | 7.9 | 4.8 | **36** | Nebraska | 3.6 | 2.9 |
| **13** | New Mexico | 7.8 | 4.8 | **37** | Iowa | 3.5 | 1.9 |
| **14** | Michigan | 7.6 | 5.4 | **38** | Kansas | 3.4 | 3.1 |
| **15** | Indiana | 7.5 | 5 | **39** | Minnesota | 3.4 | 1.6 |
| **16** | Delaware | 7.4 | 5.8 | **40** | Wyoming | 3.1 | 2.7 |
| **17** | Oklahoma | 7.4 | 4.5 | **41** | Utah | 3.1 | 2.3 |
| **18** | Kentucky | 7.2 | 3.6 | **42** | Rhode Island | 3 | 2.4 |
| **19** | Ohio | 7 | 4 | **43** | Oregon | 2.9 | 2 |
| **20** | Arizona | 6.9 | 4.7 | **44** | Massachusetts | 2.3 | 2 |
| **21** | Texas | 6.6 | 4.4 | **45** | Idaho | 2.2 | 2 |
| **22** | West Virginia | 6.6 | 4 | **46** | Vermont | 2.2 | 1.6 |
| **23** | Virginia | 6.1 | 4.1 | **47** | Maine | 1.6 | 1.6 |
| **24** | Florida | 5.9 | 5.8 | **48** | New Hampshire | 0.9 | 0.9 |

*States are sorted by intentional homicide.

**Table B7.** U.S. states by the rate of rape and aggravated assault (Federal Bureau of Investigation, 2014)

| Rank | States* | Rape (revised) | Rape (legacy) | Aggravated assault | Rank | States | Rape (revised) | Rape (legacy) | Aggravated assault |
|---|---|---|---|---|---|---|---|---|---|
| **1** | New Mexico | 70.7 | 51.4 | 421.8 | **25** | Minnesota | 36.7 | 26.6 | 123.3 |
| **2** | Michigan | 63.3 | 40.9 | 277.7 | **26** | Oregon | 36.7 | 26.5 | 140.8 |
| **3** | Arkansas | 59.4 | 39.8 | 346 | **27** | Maine | 36.5 | 27.1 | 66.9 |
| **4** | Colorado | 56.7 | 39.6 | 192.8 | **28** | Iowa | 36.3 | 26.6 | 201.6 |
| **5** | South Dakota | 55.1 | 48.4 | 245.7 | **29** | Mississippi | 35.3 | 25.5 | 153.4 |
| **6** | Montana | 52.9 | 42 | 247.4 | **30** | Rhode Island | 34.2 | 24 | 132.5 |
| **7** | Arizona | 50.2 | 36.6 | 252.1 | **31** | Indiana | 33.1 | 24.5 | 222.6 |
| **8** | Utah | 49.4 | 32.1 | 119.4 | **32** | Kentucky | 32.6 | 20 | 99.8 |
| **9** | Kansas | 48.6 | 37 | 250 | **33** | Illinois | 32.3 | 23.9 | 213.7 |
| **10** | North Dakota | 48.4 | 37.3 | 190.3 | **34** | Massachusetts | 32.3 | 24.1 | 267.6 |
| **11** | Nevada | 47.8 | 35 | 372.1 | **35** | Georgia | 30.2 | 21.4 | 218.4 |
| **12** | Nebraska | 45.8 | 33 | 176.2 | **36** | Pennsylvania | 29.9 | 21.8 | 173.5 |
| **13** | Oklahoma | 45.8 | 32.9 | 277 | **37** | Wyoming | 29.8 | 21.6 | 153.9 |
| **14** | New Hampshire | 44.8 | 34.5 | 110 | **38** | California | 29.7 | 21.6 | 236.6 |
| **15** | Ohio | 43.5 | 35.3 | 127.4 | **39** | Louisiana | 29.6 | 21.3 | 352.4 |
| **16** | Florida | 43 | 30.4 | 366.4 | **40** | Wisconsin | 29.1 | 20.3 | 170.4 |
| **17** | South Carolina | 42.8 | 30.6 | 365.8 | **41** | Virginia | 27.7 | 17.2 | 112.9 |
| **18** | Texas | 42.3 | 30.6 | 243.6 | **42** | New York | 27.5 | 19.8 | 229.4 |
| **19** | Alabama | 41.3 | 29.6 | 283.4 | **43** | West Virginia | 27.3 | 15.4 | 235.5 |
| **20** | Delaware | 41.3 | 26.6 | 306.4 | **44** | Maryland | 27.1 | 19.1 | 253.2 |
| **21** | Missouri | 39.2 | 28.1 | 304.8 | **45** | North Carolina | 24.3 | 17.5 | 215.4 |
| **22** | Tennessee | 38.6 | 28.4 | 453.2 | **46** | Connecticut | 21.7 | 15.9 | 125 |
| **23** | Washington | 38.2 | 30.7 | 164.7 | **47** | Vermont | 17.6 | 15.8 | 68.9 |
| **24** | Idaho | 37.3 | 28.6 | 160.5 | **48** | New Jersey | 14.3 | 10.7 | 125.6 |

* The states are arranged in rape (revised) order.

**Table B8.** U.S. states by the rate of divorced people and yearly divorce rate (U.S. Census Bureau, 2015; National Vital Statistics System, 2021)

| Rank | States* | Divorced percentage | Divorce rate | Rank | States | Divorced percentage | Divorce rate |
|---|---|---|---|---|---|---|---|
| **1** | Arkansas | 7.1 | 3.6 | **25** | Illinois | 5.8 | 1.3 |
| **2** | Alabama | 7.1 | 3.6 | **26** | New York | 5.8 | 2.2 |
| **3** | Florida | 7 | 3.4 | **27** | Nebraska | 5.7 | 2.6 |
| **4** | Mississippi | 7 | 3.3 | **28** | Wisconsin | 5.7 | 2.1 |
| **5** | Louisiana | 6.8 | 2.2 | **29** | Connecticut | 5.7 | 2.5 |
| **6** | Pennsylvania | 6.8 | 2.4 | **30** | Arizona | 5.6 | 2.7 |
| **7** | Oklahoma | 6.7 | 3.8 | **31** | Vermont | 5.6 | 2.3 |
| **8** | South Carolina | 6.7 | 2.4 | **32** | New Hampshire | 5.6 | 2.6 |
| **9** | Kentucky | 6.6 | 3.3 | **33** | Massachusetts | 5.6 | 1 |
| **10** | Missouri | 6.5 | 2.9 | **34** | Maryland | 5.6 | 1.6 |
| **11** | Tennessee | 6.5 | 3.3 | **35** | Nevada | 5.5 | 4.2 |
| **12** | Maine | 6.5 | 2.7 | **36** | North Dakota | 5.5 | 2.9 |
| **13** | Ohio | 6.4 | 2.6 | **37** | Wyoming | 5.5 | 3.7 |
| **14** | Delaware | 6.3 | 2.6 | **38** | Virginia | 5.5 | 3.1 |
| **15** | Iowa | 6.3 | 2.3 | **39** | Georgia | 5.5 | 2.2 |
| **16** | Rhode Island | 6.2 | 2.7 | **40** | Oregon | 5.3 | 2.7 |
| **17** | North Carolina | 6.2 | 3.2 | **41** | Minnesota | 5.2 | No data |
| **18** | New Jersey | 6.2 | 2.2 | **42** | Texas | 5.1 | 1.4 |
| **19** | Montana | 6.1 | 2.5 | **43** | California | 5 | No data |
| **20** | Indiana | 6.1 | No data | **44** | Washington | 4.8 | 2.9 |
| **21** | Michigan | 6 | 2.3 | **45** | Idaho | 4.8 | 3.4 |
| **22** | South Dakota | 6 | 2.5 | **46** | Colorado | 4.3 | 3 |
| **23** | New Mexico | 6 | No data | **47** | Utah | 3.8 | 3.3 |
| **24** | Kansas | 6 | 1.9 | **-** | West Virginia | No data | 2.9 |

* States are sorted by divorced percentage.